\documentclass[sigconf, nonacm]{acmart}
\usepackage{graphicx}
\usepackage{balance}  
\usepackage{xcolor,xspace}
\usepackage[linesnumbered,ruled,vlined]{algorithm2e}
\usepackage{amsmath,amsfonts,amsthm}
\usepackage{url}
\usepackage{bm}
\usepackage{enumerate}
\usepackage{afterpage}
\usepackage{caption}
\usepackage{subcaption}
\usepackage{booktabs}
\usepackage{hyperref}
\usepackage{enumitem}
\usepackage{nccmath}
\usepackage{pifont}

\newtheorem{definition}{Definition}

\newcommand{\eg}{\emph{e.g.}, }
\newcommand{\ie}{\emph{i.e.}, }

\newcommand\figref[1]{Fig.~\ref{#1}}

\newcommand\equref[1]{Eq.~(\ref{#1})}

\newcommand{\fakeparagraph}[1]{\vspace{1mm}\noindent\textbf{#1.}}

\newcommand{\sysname}{{\textsf{MemLens} }}

\ifodd 0

\else

\fi

\begin{document}

\title{\textsf{MemLens}: A Value-Aware Memory Management System \\with Interactive Analytics for LLM-based Agents}




\author{Shuyue Wei}
\affiliation{%
  \institution{
  C-FAIR \& School of Software, \\
  Shandong University}
}
\email{weishuyue@sdu.edu.cn}

\author{Chang Liu}
\affiliation{%
  \institution{SKLCCSE Lab, \\ 
  Beihang University}
}
\email{bjliuchang@buaa.edu.cn}


\author{Zimu Zhou}
\affiliation{%
  \institution{Department of Data Science,\\
  City University of Hong Kong}
}
\email{zimuzhou@cityu.edu.hk}


\author{Yongxin Tong}
\affiliation{%
  \institution{SKLCCSE Lab, \\ 
  Beihang University}
}
\email{yxtong@buaa.edu.cn}

\author{Lizhen Cui}
\affiliation{%
  \institution{
  C-FAIR \& School of Software, \\
  Shandong University}
}
\email{clz@sdu.edu.cn}
\begin{abstract}
Recently, memory management has become a key infrastructure for LLM-based agents, as it directly affects long-horizon reasoning, personalized responses, and knowledge reuse. 
However, existing LLM memory systems typically adopt a coarse-grained (utility-agnostic or heuristic utility) manner that treats heterogeneous user-LLM interaction records uniformly, leading to redundant and low-impact records persisting in the memory repository.
To address this challenge, we present \textsf{MemLens}, a value-aware memory management system that takes memory records as first-class data objects. 
\sysname provides an end-to-end interactive analytics dashboard that exposes the complete memory lifecycle, including Shapley-style memory evaluation, value-aware storage, and memory-assisted response. 
Through a study-copilot application, the system enables users to inspect memory values, visualize hierarchical memory structures, and compare various memory management strategies in terms of response quality, retrieval latency, and token consumption. 
Therefore, our \sysname can serve as an efficient, interpretable, and personalized long-term memory management system for agents.

\end{abstract}

\maketitle


The source code have been made available at \\  https://github.com/LIUHA1ZHU/MemLens

\section{Introduction}

    

    

\begin{figure}[thb]
    \centering
    \includegraphics[width=0.9\linewidth]{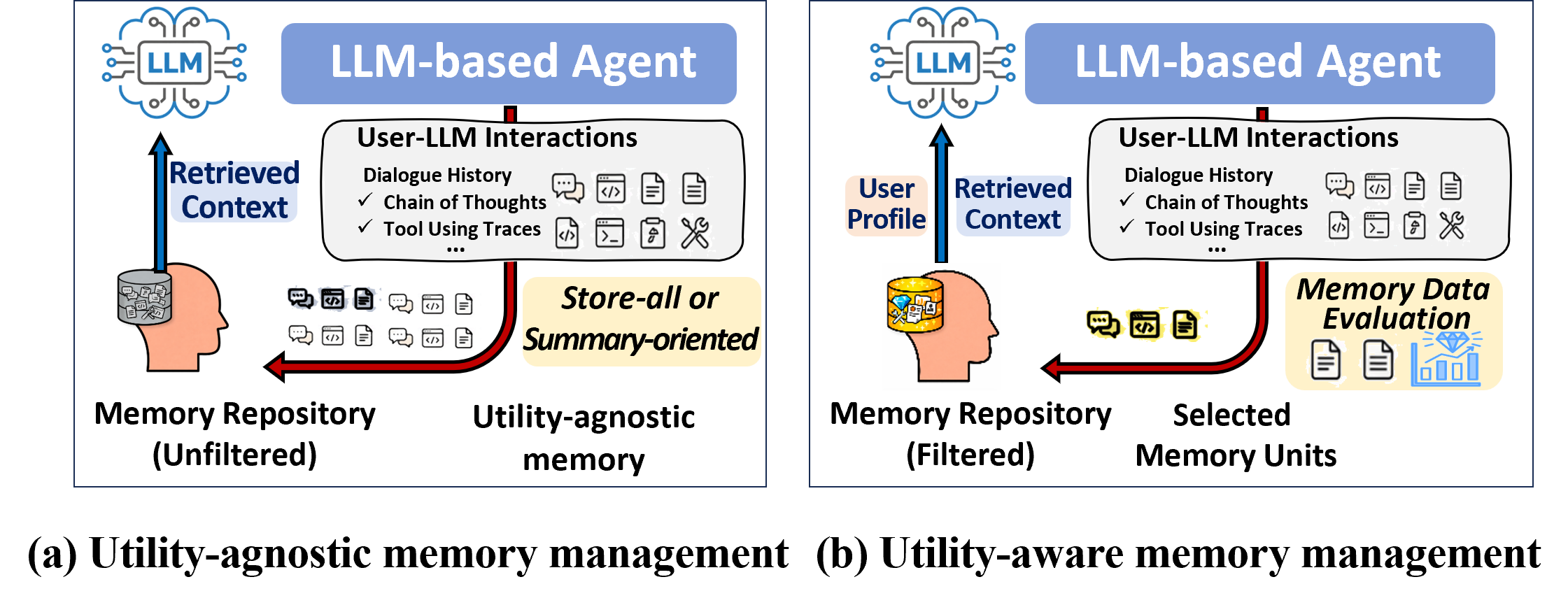}
    \caption{Comparison between (a) \textit{utility-agnostic} memory management and (b) \textit{value-aware} memory management.} 
    \label{fig:motivation}
\end{figure}

Memory data management has emerged as a key infrastructure layer for LLM-based agents, enabling long-horizon task execution, personalized interactions, and cross-session knowledge reuse.
From a data management perspective, agent memory is inherently \textit{heterogeneous}.
Beyond dialogue histories, reusable memory may also arise from intermediate reasoning traces, retrieved documents, and tool execution outputs.
{As agents operate over longer sessions, records accumulate into a growing repository of reusable context, making effective memory management increasingly important.}

However, existing LLM memory systems~\cite{TIST25_MemSurvey, arXiv25_MemSurvey, ECAI25_Mem0, arXiv26_Mem} typically manage memory in a coarse-grained manner.
They either \textit{retain nearly all} candidate records {or compress them using \textit{summary-oriented} heuristics, without explicitly modeling which memory units are truly useful for future tasks.}
{As memory accumulates, the {utility-agnostic} or {heuristic scoring} strategy allows redundant and low-impact records to persist in the repository, which can dilute task-relevant context, enlarge the retrieval space, and consume unnecessary token overhead during response generation (see \figref{fig:motivation}).}

{The core challenge is that memory usefulness is \textit{neither uniform nor static}.
A memory unit that is beneficial for one future query may be irrelevant for another, and its contribution may also depend on which other memory units are already available.
Thus, effective memory management requires more than generic summarization.
It requires a principled way to \textit{estimate the downstream utility} of heterogeneous memory units, while keeping such estimation \textit{efficient} enough to support practical storage and retrieval decisions.}

{To address this challenge, we present \textsf{MemLens}, a value-aware memory management system with interactive analytics for LLM-based agents.
\sysname manages memory based on its \textit{value}, \ie contribution of a memory unit to downstream response quality.
Specifically, it decomposes archived interactions into fine-grained memory units, estimates their utility for downstream reasoning using a lightweight proxy model together with LLM-based scoring, selectively retains high-value memory, and injects concise yet effective memory into LLM context at response time.
As the usefulness of a memory unit may depend on which other memory units are already available, \sysname models memory value through Shapley-style marginal contribution analysis, and makes this valuation practical via an efficient sampling-based approximation.
These designs yield a unified memory lifecycle spanning \textit{memory evaluation}, \textit{value-aware storage}, and \textit{memory-assisted response}.}

We demonstrate \sysname through a study-copilot application, where users can interactively explore the full memory lifecycle.
Specifically, users can:
\textit{(i)} inspect how archived user-LLM interactions are decomposed into memory units and assigned value-aware scores;
\textit{(ii)} visualize how the memory evolves as retention is guided by value thresholds and hierarchical organization; and
\textit{(iii)} compare various memory management strategies, including \textit{store-all}, agent-based summarization, and proposed \textit{value-aware} strategy, in terms of response quality, retrieval latency, and token consumption.

\fakeparagraph{Contributions}
The main contributions are summarized as follows:
\textit{(i)} 
We introduce a value-aware memory management paradigm for LLM-based agents that explicitly models memory value as downstream utility, enabling selective retention of high-impact memory units.
\textit{(ii)} 
We design an end-to-end memory lifecycle framework that integrates memory valuation, value-aware storage, and memory-assisted response, providing a unified and interpretable mechanism for long-term memory management in LLM-based agents.
\textit{(iii)} We develop an interactive demonstration system that visualizes memory value, storage evolution, and response behavior, allowing users to directly observe how different memory strategies affect response quality, retrieval latency, and token consumption.

\begin{figure}[htb]
    \centering
    \includegraphics[width=0.98\linewidth]{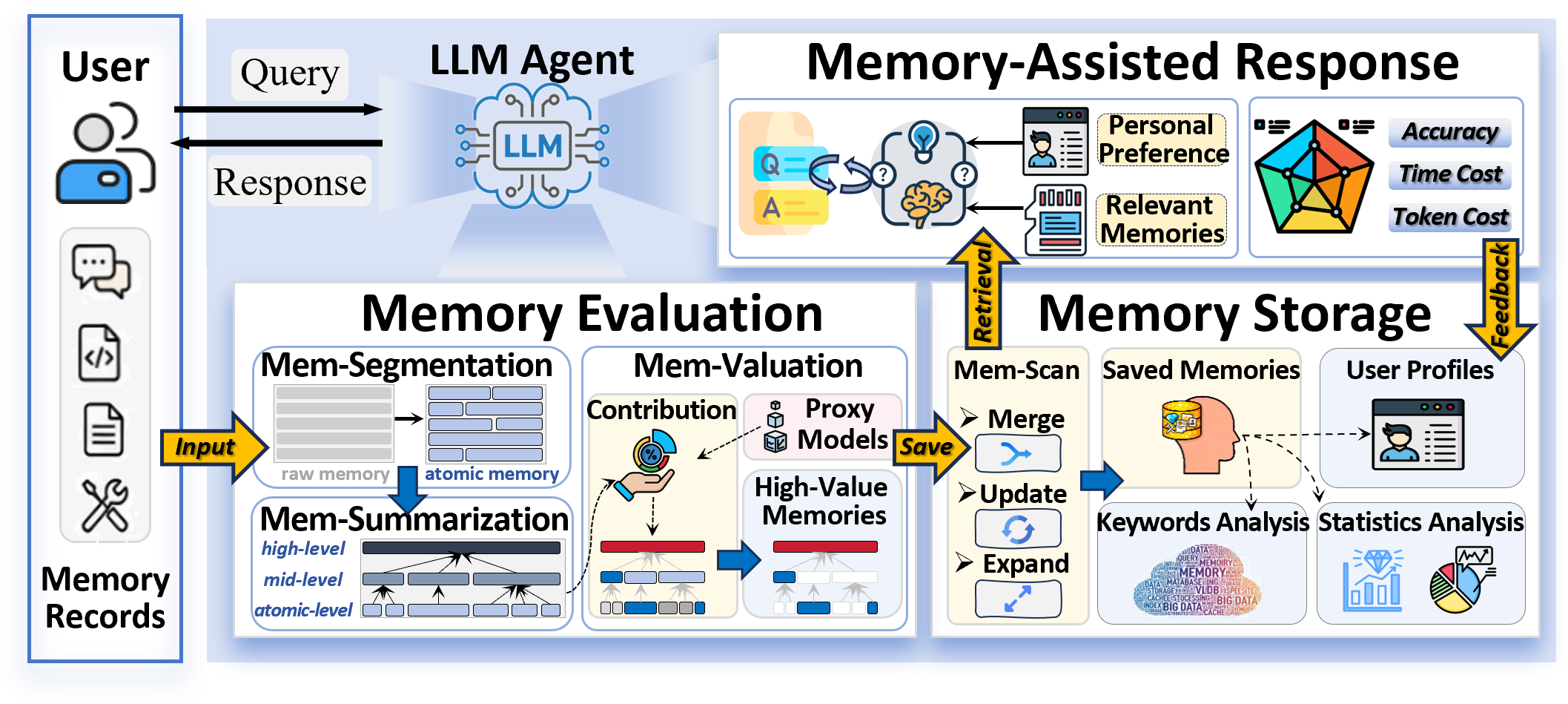}
    \caption{System Overview of \sysname}
    \label{fig:overview}
\end{figure}

\section{\textsf{\sysname} Overview} \label{sec:overview}




\fakeparagraph{Overview} \sysname is a value-aware memory management system that shifts how LLM-based agents store and utilize long-term memory. 
As illustrated in \figref{fig:overview}, it integrates three core components, including memory data evaluation, value-aware memory storage, and memory-assisted response.
The system treats heterogeneous records originating from the LLM-based agent’s reasoning process (such as dialogue histories, retrieved knowledge, and chain-of-thought traces) as candidate memory records. 
Instead of indiscriminately storing all records, \sysname explicitly quantifies their utility and selectively retains only the high-value ones.
During query response, the system retrieves and injects the concise but effective memories to support complex reasoning, enabling more accurate responses with significantly lower token overhead.

Furthermore, \sysname also features an interactive analytics dashboard that makes value-aware memory management transparent and visible. 
Users can inspect memory values on \sysname, where each record is assigned a utility score and aggregated into global views such as value distributions and retention ratios, enabling identification of the critical versus redundant memories.  
The system also presents an interpretable storage view that highlights salient keywords and evolving user preference profiles, revealing how long-term memory structures are continuously shaped. 
In addition, \sysname enables side-by-side comparison of different memory management strategies (\eg store-all, agent-based summarization, and value-aware storage), reporting response quality, retrieval latency, and token consumption on testing datasets, which allows users to directly observe how value-aware memory reduces unnecessary context while enhancing the response quality.

\fakeparagraph{Workflow} \sysname implements a value-aware memory lifecycle that enables LLM agents to retain and utilize the long-term memory. 
Once the user archives a session, the system can decompose raw records into fine-grained memory units and organize them into a hierarchical structure with multiple levels of abstraction.  
Then, each memory unit is assigned a value using a widely-adopted Shapley-value-based framework~\cite{FCS26_Shapley, FCS25_FedLLM_Survey} that captures its contribution to downstream reasoning performance. 
Guided by these values, \sysname selectively preserves high-value memories and continuously updates a user preference profile.  
Finally, the system responds to the user query by retrieving relevant memory units and using the user profile, then injecting them into the LLM context to enhance response quality while avoiding unnecessary context expansion. 
The above closed-loop lifecycle allows \sysname to maintain compact yet high-impact and personalized memory records adaptively.

\section{\textsf{\sysname} System Design} \label{sec:designs}


\subsection{Memory Data Evaluation}

\sysname treats memory records as first-class data objects and quantifies their value via a principled Shapley-value-based approach in two steps: \textit{memory unit construction} and \textit{value estimation}.

\fakeparagraph{Memory Unit Construction}
\sysname converts raw user-LLM interaction records into structured memory units.
Firstly, for fine-grained analysis, the raw record is segmented into sentence-level atomic units.
The system then groups the memory units into multi-level summaries via LLMs, providing a hierarchical memory structure that captures both detailed semantics and high-level knowledge. 
We maintain provenance links from summarized nodes to the atomic memory units as well
to avoid the information loss.
This design allows the \sysname system to replace multiple low-utility fragments into a single high-value alternative, which potentially improves both storage efficiency and memory quality.

\noindent\textbf{Memory Value Estimation.}
\sysname assesses the utility of each memory unit through an LLM-as-Judge utility score, which is a different judge model from the one employed during memory valuatio and measures unit contribution to the downstream reasoning task. 
Intuitively, a memory unit is valuable if it can significantly improve the output quality of a lightweight proxy model $\medop{\mathcal{M}_{p}}$ with limited reasoning capability.
By default, MemLens uses Qwen2.5-7B as the proxy model in EduMemBench, while the dashboard allows users to switch among different lightweight models.
Specifically, given the original user query $q$ and its related memory units subset $S \subseteq \textsf{M}=\{m_1, m_2, \cdots, m_M\}$, we can formalize the utility function $\mathcal{V}(\cdot)$ of a unit set $S$ as follows,
\begin{equation} \label{eq:utility}
\mathcal{V}_{q}(S) = \textbf{Score}\big(\mathcal{M}_{p}(q \oplus S)\big) - \textbf{Score}\big(\mathcal{M}_{p}(q)\big),
\end{equation}
where $\mathcal{M}_{p}$ is the proxy model, $\oplus$ denotes the prompts injection operation, and $\textbf{Score}(\cdot)$ can be implemented via {LLM-as-Judge} approaches. 
As \equref{eq:utility} enables \sysname to qualify the utility of a set of memory units $S$, we can define the value of a memory unit $m_i \in \textsf{M}$ by extending the widely-adopted Shapley value as follows.
\begin{definition}[Memory Shapley Value] Given a user query task $q$ and its related memory units $\textsf{M}=\{m_1,\dots,m_{M}\}$ from the user-LLM interactions, the memory value of $m_i$ can be formally defined as,
\begin{equation}
    \phi(m_i) = \sum_{S \subseteq \textsf{M} \setminus \{m_i\}} \frac{|S|!(M - |S| - 1)!}{M!} \left[ \mathcal{V}_{q}(S \cup \{m_i\}) - \mathcal{V}_{q}(S) \right],
\end{equation}
    where $M$ is the total number of memory units.
    The \textit{Memory Shapley Value} (MS-value) extends the Shapley value-based data valuation concepts and inherits some required fairness properties, such as null player (or no-free-riders), symmetric fairness, and linear-additivity.
\end{definition}

To support in-time interactive memory data analytics and feedback, we further extend a stratified sampling strategy~\cite{ICDE25_FedShapley} to approximate memory values efficiently, where subsets $S$ are grouped
by size, and samples are drawn proportionally to reduce approximation errors.
Then, the evaluation time complexity can be reduced to $O(\rho M)$, where $\rho$ is the sampling budget and $\rho$ is usually set to $20\sim100$ for efficient interactive latency.
In summary, this module enables us to save high-utility memories selectively and provides transparent and fine-grained insight into the value of memory units.

\subsection{Value-Aware Memory Storage} 
Building upon the computed Memory Shapley values, this module implements a value-aware memory storage that handles memory retention, organization, and evolution under explicit utility constraints, providing a compact yet high-utility memory warehouse that balances storage cost and reasoning performance. 

\fakeparagraph{Value-Aware Storage Strategy} For memory units $\textsf{M}$ and their associated with values $\{\phi(m_i)\}$, \sysname can make selectively storage by a value threshold $\tau$, $
\textsf{M}^{*} = \{ m_i \in \textsf{M} \mid \phi(m_i) \ge \tau \}$, allowing users to flexibly adjust $\tau$ and explore a trade-off between storage cost and downstream reasoning performance.
We also achieve interactive threshold tuning to provide immediate feedback on memory size, compression rate, and expected utility in \sysname system.

\fakeparagraph{Memory Organization and Consolidation} 
In \sysname, the saved memory units are organized into a hierarchical tree derived from previous multi-level abstractions.
As sentence-level atomic memory units have been grouped into higher-level semantic nodes, they can form a memory tree in which leaf nodes are atomic memory units and internal nodes represent their high-level abstractions.
To reduce memory redundancy, \sysname will incrementally add a new memory unit to the existing memory tree.
Specifically, the system applies an LLM-aided consolidation operator to decide how to handle $m{'}$:
\textit{(i)} merge $m{'}$ into an existing node, \textit{(ii)} update an existing node in memory tree, or \textit{(iii)} insert $m'$ as a new node.
The operator first computes semantic similarity between the incoming memory unit and existing nodes.
Highly similar nodes are merged, partially overlapping nodes are updated, and unrelated units are inserted as new nodes.
The memory tree is also visualized in \sysname, allowing users to inspect memory organization and semantic relationships directly. 

\fakeparagraph{Interactive Memory Analytics} 
Our demonstration system also provides an interactive analytics interface to improve transparency and interpretability of the memory lifecycle. 
Specifically, the user profiles are constructed as a structured summary of retained high-value memory units and injected as system prompts, which provides a \textit{word-cloud-based exploration} of saved memory content, and presents \textit{key memory statistics}, including value distributions, memory size, and compression rates. 
These allow users to observe how a value-aware storage strategy affects the saved LLM memory units.

\subsection{Memory-Assisted Response}

This module uses value-aware memories and user profiles to generate personalized responses, then allows users to archive completed sessions for continuous memory evolution and management.

\fakeparagraph{Response Generation}
\sysname first uses the user preference profile, constructed from historical high-value memories, as the \textit{system prompts} to maintain consistent personalization across sessions. 
Then, the system retrieves relevant memories from the integrated vector database and re-ranks the retrieved units based on their MS-value $\phi(m)$, prioritizing units that are both semantically relevant and empirically high-utility. 
These selected memory units are injected into the LLM prompts to enhance response quality.

\fakeparagraph{Memory Archiving}
Finally, the user can archive the interaction records and feed them into the value-aware memory pipeline for memory unit construction, value estimation, and selective storage, which  establishes a closed-loop LLM memory lifecycle that continuously adapts to evolving user behavior and query requirements. 
\section{Demonstration Scenarios}

We demonstrate \sysname through an interactive study-copilot scenario, where it assists students in understanding key concepts, creating profiles, and maintaining memory across sessions. 

\fakeparagraph{Simulation Environment}
To emulate realistic user behavior, we construct a synthetic benchmark dataset, \textsf{EduMemBench}, containing over 2,000 multi-turn dialogue sessions generated by an LLM-based student agent, covering core database topics such as transactions, indexing, and concurrency control. 
Based on these interactions, we further derive a testing set of 500 query--answer pairs with reference answers, enabling systematic evaluation of memory-assisted response quality, retrieval latency and token consumption overhead under different memory management strategies.

\fakeparagraph{Interactive Demonstration Workflow}
The \sysname provides three coordinated interfaces corresponding to the LLM memory lifecycle: \textit{(i) memory evaluation}, \textit{(ii) value-aware memory storage}, and \textit{(iii) memory-assisted response}. 
Through \sysname users can interactively explore how LLM memory units are valued, organized, and utilized, which creates a complete end-to-end experience.

\begin{figure}[thb]
    \centering
    \includegraphics[width=0.98\linewidth]{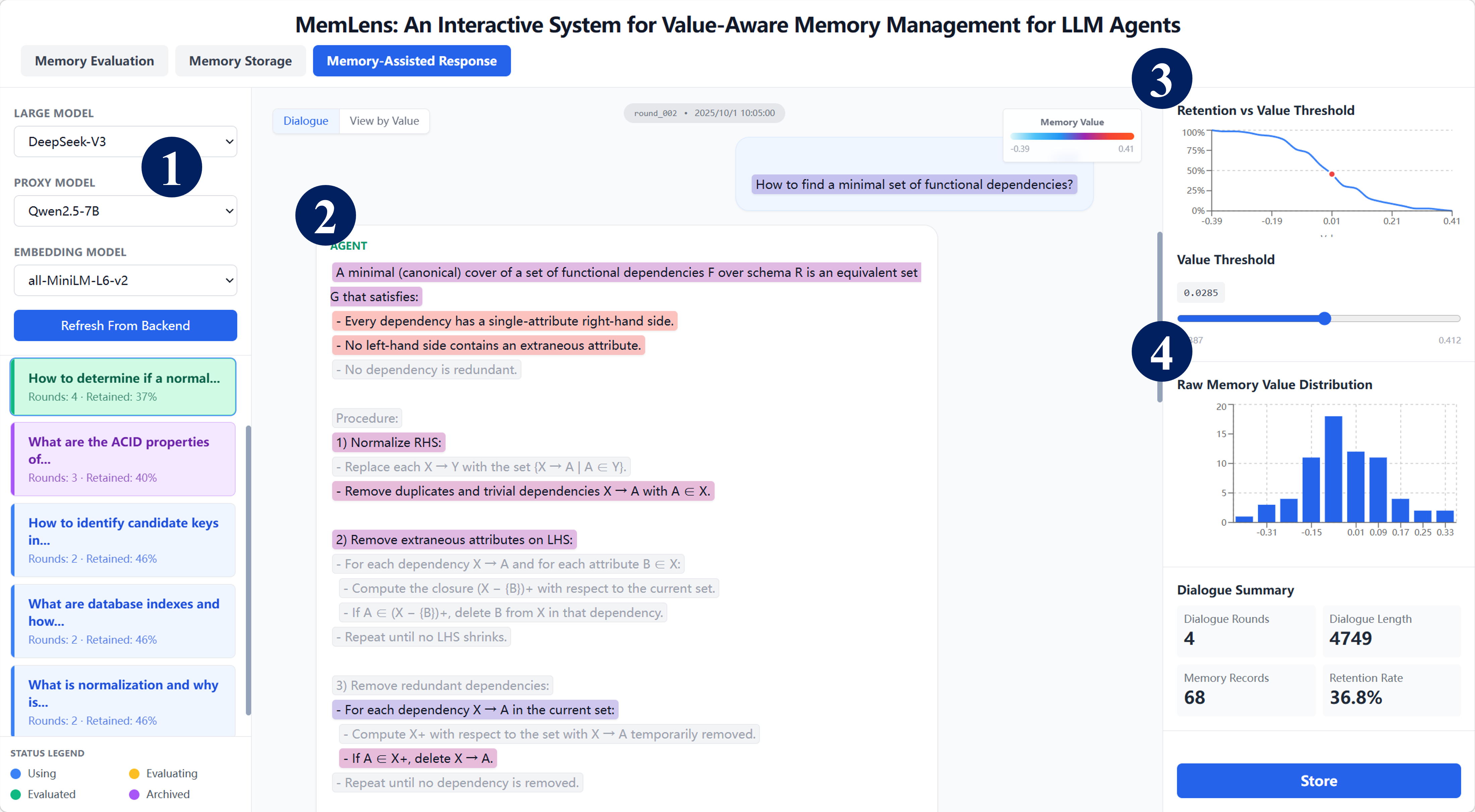}
    \caption{Memory Evaluation Interface}
    \label{fig:ifc1}
\end{figure}

\noindent\textit{(i) Memory Evaluation Interface} (in \figref{fig:ifc1}).
Users can first explore how memory values are assigned. 
\ding{172} The system supports configurable evaluation settings, allowing users to select different LLMs as scoring models and lightweight proxy models for utility estimation. 
\ding{173} Each archived session is decomposed into fine-grained memory units, whose values are visualized via color-coded highlights, enabling immediate identification of high- and low-value memory fragments. 
\ding{174} An interactive threshold controller allows users to dynamically adjust the retention threshold and observe corresponding changes in memory compression. 
\ding{175} The memory evaluation interface further presents value distributions, revealing the empirical patterns of memory across multiple sessions.

\begin{figure}[thb]
    \centering
    \includegraphics[width=0.98\linewidth]{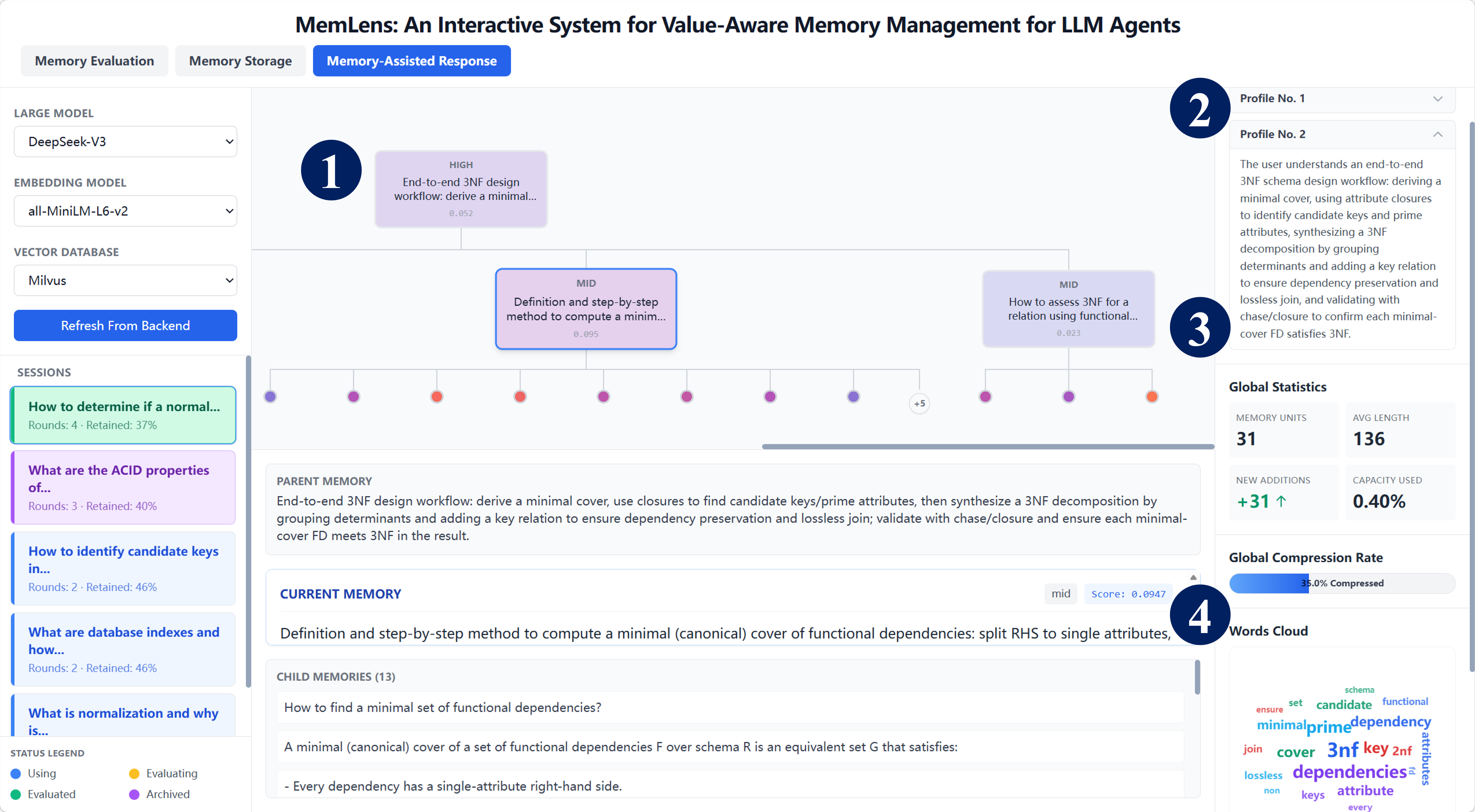}
    \caption{Value-Aware Memory Storage Interface}
    \label{fig:ifc2}
\end{figure}

\noindent\textit{(ii) Value-Aware Memory Storage Interface} (in \figref{fig:ifc2}).
Users then inspect how memory is organized and stored. 
\ding{172} The system visualizes the hierarchical memory structure as an interactive tree, where node colors encode memory utility. 
Users can explore the tree by clicking nodes in the memory tree to examine their contents and semantic relationships. 
\ding{173} This panel continuously updates a user preference profile derived from retained memories, reflecting long-term learning behavior. 
\ding{174} This panel presents memory statistics (\eg memory unit size, compression rate, and storage capacity), enabling users to observe how value-aware storage reshapes the memory repository.
\ding{175} \sysname also provides a word-cloud view that highlights the key concepts in stored memory based on their frequency, enabling users to quickly grasp the dominant memory patterns during their personalized knowledge learning process.

\begin{figure}[thb]
    \centering
    \includegraphics[width=0.98\linewidth]{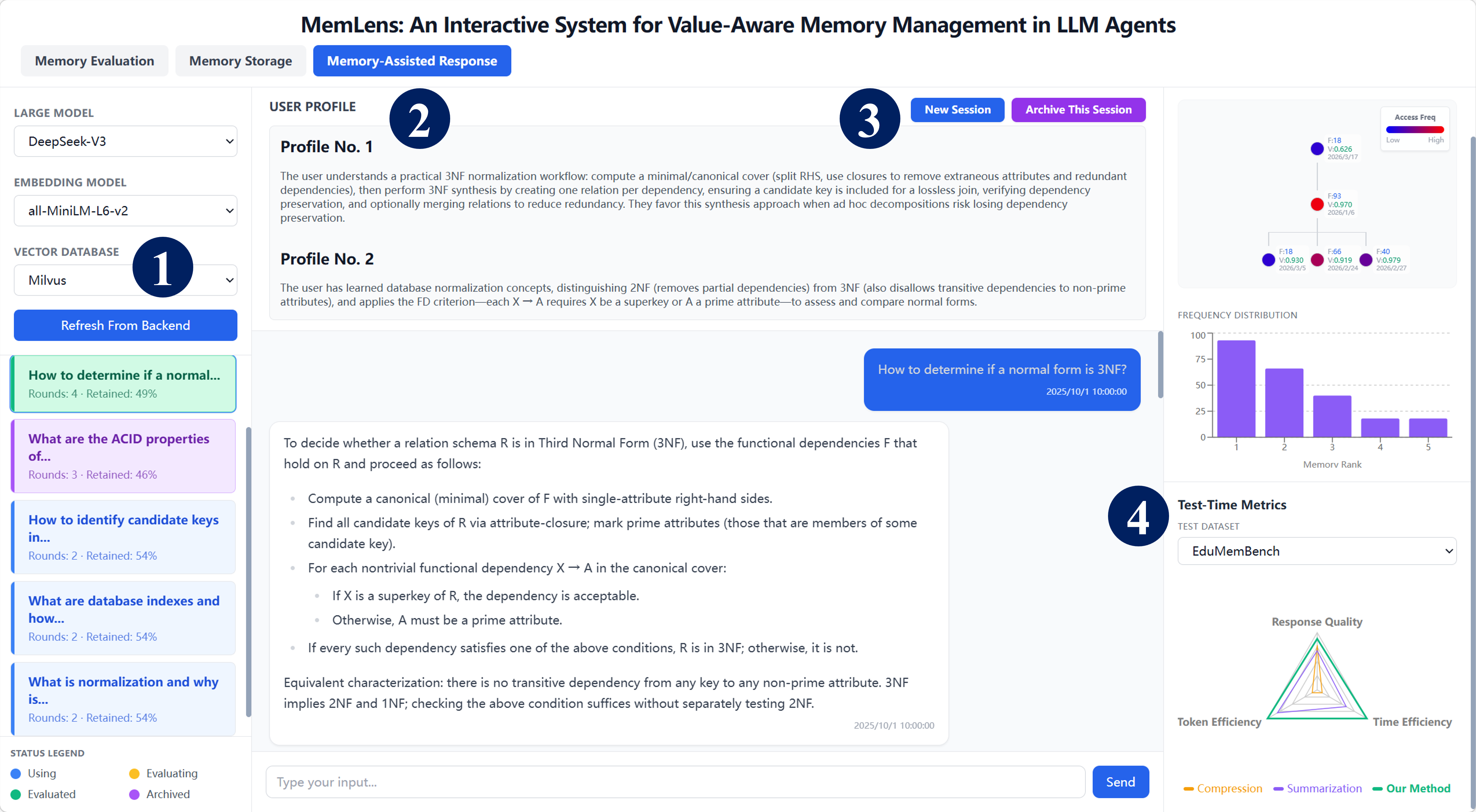}
    \caption{Memory-Assisted Response Interface}
    \label{fig:ifc3}
\end{figure}

\noindent\textit{(iii) Memory-Assisted Response Interface.} (in \figref{fig:ifc3}).
Finally, users interact with the system in real time to observe the impact of memory on responses, which helps them to recall previous interactions. 
\ding{172}Users can select different vector backends (\eg FAISS~\cite{TBD25_FAISS}) to set system configurations. 
\ding{173}The system generates responses using both retrieved memory and user profiles, allowing users to observe improvements in answer relevance and consistency directly. 
\ding{174}Users can archive completed sessions, triggering the full memory pipeline, thereby closing the memory lifecycle. 
\ding{175}This panel supports benchmarking across different memory strategies and also presents scores of response quality, retrieval time, and token consumption in a radar plot, enabling users to observe the trade-offs and advantages of various memory strategies.
Experiments on {EduMemBench} shows that the value-aware strategy consistently improves the trade-off among response quality, latency, and token consumption.
Though EduMemBench is used as the primary demonstration workload, MemLens is benchmark-agnostic and can be readily extended to public long-term memory benchmarks such as LoCoMo\cite{LOCOMO_Data} and LongMemEval\cite{LMES_Data}.

\bibliographystyle{ACM-Reference-Format}
\bibliography{ref}

@article{TIST25_MemSurvey,
  author       = {Zeyu Zhang and
                  Quanyu Dai and
                  others},
  title        = {A Survey on the Memory Mechanism of Large Language Model-based Agents},
  journal      = {{ACM} TOIS},
  volume       = {43},
  number       = {6},
  pages        = {155:1--155:47},
  year         = {2025},
}

@article{arXiv25_MemSurvey,
  title={Memory in the age of ai agents},
  author={Hu, Yuyang and Liu, Shichun and Yue, Yanwei and Zhang, Guibin and Liu, Boyang and others},
  title        = {Memory in the Age of {AI} Agents},
  journal      = {arXiv},
  volume       = {abs/2512.13564},
  year         = {2025},
}

@inproceedings{ECAI25_Mem0,
  author       = {Prateek Chhikara and
                  Dev Khant and
                  Saket Aryan and
                  others},
  title        = {Mem0: Building Production-Ready {AI} Agents with Scalable Long-Term Memory},
  booktitle    = {ECAI},
  volume       = {413},
  pages        = {2993--3000},
  publisher    = {{IOS} Press},
  year         = {2025},
}

@article{FCS26_Shapley,
      author       = {Yuxiang Wang and
                      Shuyuan Li and
                      others},
      title        = {Efficient shapley-based data valuation for federated trajectories},
      journal      = {Frontiers Comput. Sci.},
      volume       = {20},
      number       = {9},
      pages        = {2009621},
      year         = {2026},
}

@inproceedings{ICDE25_FedShapley,
  author       = {Shuyue Wei and
                  Yongxin Tong and
                  Zimu Zhou and
                  Tianran He and Yi Xu},
  title        = {Efficient Data Valuation Approximation in Federated Learning: {A}
                  Sampling-Based Approach},
  booktitle    = {ICDE},
  pages        = {2922--2934},
  publisher    = {{IEEE}},
  year         = {2025},
}

@article{TBD25_FAISS,
  title={The faiss library},
  author={Douze, Matthijs and Guzhva, Alexandr and Deng, Chengqi and Johnson, Jeff and Szilvasy, Gergely and others},
  journal={IEEE Transactions on Big Data},
  year={2025},
  publisher={IEEE}
}

@inproceedings{LMES_Data,
  author       = {Di Wu and
                  Hongwei Wang and
                  Wenhao Yu and
                  others},
  title        = {LongMemEval: Benchmarking Chat Assistants on Long-Term Interactive
                  Memory},
  booktitle    = {ICLR},
  publisher    = {OpenReview.net},
  year         = {2025},
}

@inproceedings{LOCOMO_Data,
  author       = {Adyasha Maharana and
                  Dong{-}Ho Lee and
                  Sergey Tulyakov and
                  others},

  title        = {Evaluating Very Long-Term Conversational Memory of {LLM} Agents},
  booktitle    = {ACL},
  pages        = {13851--13870},
  publisher    = {Association for Computational Linguistics},
  year         = {2024},
}

@article{FCS25_FedLLM_Survey,
  author       = {Shuyue Wei and
                  Yongxin Tong and
                  Zimu Zhou and
                  others},
  title        = {Federated reasoning LLMs: a survey},
  journal      = {Frontiers Comput. Sci.},
  volume       = {19},
  number       = {12},
  pages        = {1912613},
  year         = {2025},
}

@article{arXiv26_Mem,
  author       = {Jiajie Fu and
                  Junwen Chen and
                  others},
  title        = {VikingMem: {A} Memory Base Management System for Stateful LLM-based Applications},
  journal      = {arXiv},
  volume       = {abs/2605.29640},
  year         = {2026},
}

\end{document}